\documentclass[pre,twocolumn,showpacs,amsmath,amssymb]{revtex4-1}

\usepackage{graphicx}
\usepackage{graphicx}
\usepackage{dcolumn}
\usepackage{bm}
\usepackage[T1]{fontenc}
\usepackage[usenames]{color}

\usepackage{amssymb}
\usepackage{amsmath}

\large

\begin{document}
\def\bw{\begin{widetext}}
\def\ew{\end{widetext}}
\def\dg{\dagger}
\def\ddg{\ddagger}

\preprint{APS/123-QED}
\title{On the influence of the "donor"/"acceptor" presence on the excitation states in molecular chains: non--adiabatic polaron approach}
\author{V. Matic$^a$, Z. Ivic$^{a}$. Przulj$^{a}$, D. Chevizovich$^{a}$}
\email{cevizd@vin.bg.ac.rs}
\affiliation{$^a$ Vinca Institute of Nuclear Sciences--National Institute of the Republic of Serbia, University of Belgrade, P.O. BOX 522, 11001, Belgrade, Serbia}

\date{\today}

\begin{abstract}
In the paper, we considered a molecular structure that consists of a molecular chain and an additional molecule ("donor"/"acceptor") that can inject (or remove) single excitation (vibron, electron, e.t.c.) onto the molecular chain. We assumed that the excitation forms a self--trapped state due to the interaction with mechanical oscillations of chain structure elements. We analyzed the energy spectra of the excitation and showed that its state (when it migrates to the molecular chain) has the properties of the non--adiabatic polaron state. The conditions under which the excitation can migrate from one subsystem to another were considered. It was shown that the presence of a "donor" molecule cannot significantly change the properties of the excitation located on the molecular chain. At the same time, the molecular chain can affect the position of the energy level of the excitation localized on the "donor" subsystem. Indirectly, this can influence the process of excitation migration from one subsystem to another one. The influence of basic energy parameters of the system and the environment temperature on this process are discussed. The entire system was assumed to be in thermal equilibrium with the environment.
\end{abstract}



\maketitle


\section{Introduction}

The migration of a single quantum of energy or charged particle at the submolecular level plays an important role in biological processes \cite{DavydovBQM,DavydovSMS,Petrov,Voet,Dauxois,ZdravkovicCevizovicND}. In many of them, the excitation transfers from one molecule to another one ("donor" or "acceptor" molecule) \cite{DavydovBQM,PetrovJTB, BrizhikPRE2014,ChenACIE,Frohchlich}. Sometimes, the excitation crosses large distances. Numerous models have been developed to explain such processes. It is believed that excitation migrates by the quantum tunneling mechanism when the "donor" and the "acceptor" molecules are close. But if the distance between the molecules is large, quantum tunneling is practically impossible and other mechanisms should be considered. Some of them are based on the assumption that the excitation migration between two molecules takes place through the so--called "molecular bridge". For example, the role of a "molecular bridge" has a molecular chain (MC) that physically connects the place where the excitation emerges with that where it uses in some biophysical process \cite{DavydovBQM,DavydovSMS,Voet}. During migration, excitation transfers from one subsystem to another (or from one structure element of MC to another one along the same molecular chain). Here, an important question arises: how the excitation can be transferred from one end of the molecular structure to the other one without its energy disappearing in some dissipation process? In other words, how to explain the high efficiency of excitation transfer?\\

A definitive answer to this question has not yet been offered, at least not in the form of a model that can be applied to a wide class of biomolecular systems and for different types of excitations. To illustrate more clearly what we mean, let us mention some examples. Firstly, let us mention the energy released during the hydrolysis of ATP to ADP, which is necessary for numerous physiological functions of a living cell \cite{DavydovBQM,DavydovSMS,Petrov,Voet,Dauxois, ZdravkovicCevizovicND}. This quantum of energy is not sufficient to excite an electronic state of the molecule, but it can excite the intramolecular oscillation mode (Amide--I vibron mode) on the nearest peptide bond of the polypeptide MC \cite{DavydovBQM,DavydovSMS,Petrov,Dauxois, ZdravkovicCevizovicND, CruzeiroLTP,ScottPR,PouthierPRL,CevizovicPRE,CevizovicCPB,FalvoPouthier,PuthierPRE2003,AK,PouthierPRE2008}. Due to the interaction with the mechanical (thermal) oscillations of the structural elements of the MC, the vibron becomes self--trapped (ST) \cite{AK,PekarZETF1946,PekarZETF1948,Rashba,EminPRB}. In this form, it can migrate along the entire MC practically without energy losses. Various polypeptide MCs have been considered as a medium for vibron transfer: protein molecules with an alpha--helix secondary structure, for example \cite{DavydovBQM,DavydovSMS, CevizovicCPB,PouthierPRE2008,PouthierPRL,CSF2015Ceviz}. We should have in mind that on the ends of the protein molecules, there are prosthetic groups (part of the protein structure, tightly connected to the rest of the molecule). They can influence the vibron properties and prevent or force the appearance of a ST excitation state in the MC \cite{BrizhikPRE2014,Voet}. In addition, they can affect what will happen with the excitation that comes to the end of the protein chain: whether it will be reflected or absorbed by the prosthetic group, after which it will be spent for a metabolic process. Another important example of excitation migration is the transfer of electrons (holes or other charged particles) through biomolecules. In redox processes during cellular respiration, an electron migrates from a "donor" molecule to an "acceptor" through the "molecular bridge" \cite{Voet,BrizhikPRE2014}. When it passes the "molecular bridge", the electron migrates from one structural element of the MC to the neighboring one, up to the place from which it migrates to the "acceptor" molecule. The transport of an electron along the MC also takes place during the photosynthesis process \cite{DavydovBQM,DavydovSMS,Voet,ScottPR}. As in the case of vibron migration, the presence of the "donor" and "acceptor" molecules can influence the properties of the "molecular bridge" but also the charged particle themselves.\\

The "molecular bridges" in the above--mentioned examples must provide high efficiency of excitation transfer over large distances (those comparable to the size of the "molecular bridge"). So far, various hypotheses and corresponding physical models have been put forward but none of them has provided a satisfactory explanation of the mentioned processes. The considered models usually are based on the assumption that the excitation forms a soliton wave (in the case of one--dimensional regular structures it corresponds to an adiabatic large polaron) \cite{DavydovBQM,DavydovSMS,BrizhikPRE2014,CruzeiroLTP,DavydovPSS, DavydovZETF, CSF2015Ceviz,ScottPR,IvicCP426,CSF2015Ceviz, Schuttler,GogolinPR1988,Young,Shaw, Campbell,CastroNetoCaldeira}. Such soliton has high stability and can migrate over large distances without dissipation. However, in the mid--80s some researchers expressed their doubts about soliton models and began to consider other models, especially in the case of vibron excitation self--trapping \cite{AK,PouthierPRL, FalvoPouthier,CevizovicCPB,CevizovicPRE}. Their criticism was based on the fact that biological macromolecules do not possess such physical properties required for the soliton formation in the case of vibron self--trapping. Because of that, they assumed that the non--adiabatic polaron appears due to the interaction of vibron with mechanical oscillations of the structural elements of the MC \cite{AK,FalvoPouthier,CevizovicPRE,CevizovicCPB, ZdravkovicCevizovicND}.\\

On the other hand, the structures that participate in the excitation transfer (which are either part of the MC or are separate molecular structures interacting with the MC) must not disturb the stability of the quasiparticle state formed on the MC. In the further text, such structures were named as donor/acceptor (D/A) subsystems. As a consequence, an important problem in the processes of transfer of excitation from the D/A molecule to the MC (and vice versa) is the influence of the presence of the D/A molecule on the properties of the self--trapped excitation. So far, this problem has been considered within the framework of soliton theory \cite{DavydovBQM, BrizhikPRE2014}. However, under the assumption that the excitation in MC forms a non--adiabatic polaron, the problem of the influence of the D/A molecule on self--trapped excitation has not been considered so far. The essential difference between soliton models and the non--adiabatic ppolaron model is that previously mentioned models are based on the assumption that the continuum approximation is applicable in the case of excitations that have different natures (vibron, electron, hole,...). On the other hand, the values of some of the basic energy parameters of biomolecular chains are not exactly known \cite{AK,PouthierPRE2008,CevizovicPRE,CevizovicCPB}. Even more, estimates of the values of these parameters in the case of vibron excitation in polypeptide molecules put into question the applicability of continuous approximation. In addition, one more interesting question can be asked here: what properties a molecule must have so that it could be a donor or acceptor of excitation when it comes into contact with a "molecular bridge"? Especially if we consider that the excitation on the MC forms a "dressed" particle.\\

In this paper, we consider the influence of the presence of the D/A molecule on the ST state of single excitation, injected into the molecular chain. The discussion is based on the assumption that the stability of the excitation in MC is the consequence of its self--trapping and the formation of dressed quasi--particle, with properties similar to the properties of {\bf non--adiabatic (small) polaron}. Namely, due to the interaction of single excitation with the phonons in MC (which is in thermodynamic equilibrium with its environment), the excitation forms dressed quasi--particle whose energy states are more favorable than the energy of a "bare" excitation. Although the appearance of the non--adiabatic polaron is justified only for vibron excitations \cite{AK,ZdravkovicCevizovicND}, we used this term in a more general sense.\\

Usually, the D/A molecule is much smaller and most often weakly interacts with the MC. Consequently, in many of the existing theoretical studies devoted to the excitation transport in biostructures consisting of the molecular bridge and the D/A molecule, the D/A molecules are treated as weakly connected to the "molecular bridge" \cite{DavydovBQM,BrizhikPRE2014}. Nevertheless, in this paper, we supposed that the presence of D/A molecule can (in principle) significantly affect the energy spectrum of the ST excitation in MC and the degree of its dressing. However, we neglected the interaction of the D/A molecule with the mechanical oscillations of MC. We also assumed that the influence of the presence of the D/A subsystem on the excitation is local. Consequently, the excitation only can migrate between the D/A molecule and the nearest structural element of the MC.


\section{The model}

\begin{figure}[h]
	\begin{center}
		\includegraphics[height=2cm]{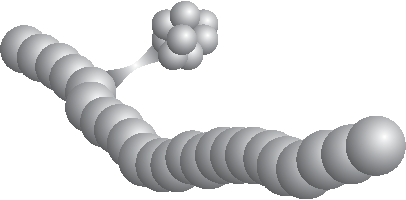}\\
		\vskip 1cm
		\includegraphics[height=2cm]{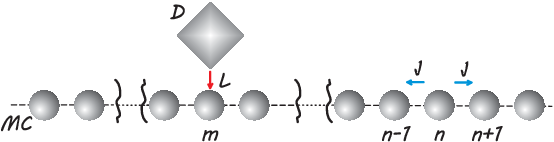}
	\end{center}
	\vskip -0.5cm
	\caption{Upper pane: schematic presentation of the D/A--MC structure. Lower pane: simplified D/A--MC structure. Although the excitation transfers between the D/A and a particular peptide group of MC, here the sphere represents both the peptide group and the rest of structure element of MC.}\label{fig1}
\end{figure}

Let us consider a single excitation (vibron, electron, etc.) injected in the structure that consists of a D/A molecule and the MC (D/A--MC system). The MC consists of $N$ structure elements (denoted with labels 1,2,...,$N$ on Fig.(\ref{fig1})), while the D/A molecule is located near some $m$--th structure element of the MC.

\noindent As a theoretical framework, we applied the Holstein molecular crystal model \cite{HolsteinAP,CevizovicPRE,CevizovicCPB,ZdravkovicCevizovicND}, modified to account for the presence of the D/A molecule:

\begin{equation}\label{PocH}
\hat{H}=\hat{H}_D+\hat{H}_{MC}+\hat{H}_{D-MC}+\hat{H}_{ph}+\hat{H}_{MC-ph}
\end{equation}

\noindent Here $$\hat{H}_D=\mathcal{E}_D\hat{D}^{\dag}\hat{D}$$ describes single excitation, excited on the D/A molecule. Operators $\hat{D}^{\dg}$ and $\hat{D}$ are creation and annihilation operators of the excitation at the state with energy $\mathcal{E}_D$. The term $$\hat{H}_{MC}=\mathcal{E}_0\sum_n \hat{B}^{\dag}_n\hat{B}_n-J\sum_n \hat{B}^{\dag}_n\left(\hat{B}_{n+1}+\hat{B}_{n-1}\right)$$ is the Hamiltonian describing the excitation located on MC. Operators $\hat{B}^{\dg}_n$ and $\hat{B}_n$ are creation and annihilation operators of the excitation, excited on the $n$--th structural element of the MC. The $\mathcal{E}_0$ is the energy required to excite the corresponding excitation mode on the particular structure element of the MC (in the case of an electron in a polypeptide MC, it is the excitation energy of the weakest bound electron in peptide group: the extra electron injected on the $n$--th peptide group is at the state determined by the energy level $\mathcal{E}_0$). The $J$ is the transfer--integral between neighboring structure elements of MC (in the vibron case it is the energy of the resonance dipole--dipole interaction; in the case of electron it corresponds to the overlap of electronic orbitals between neighboring molecules). From the physical point of view, the term with $J$ in $\hat{H}_{MC}$ describes the excitation jumps from one of the MC nodes to the neighboring one due to the resonant dipole--dipole interaction (or the overlap of the corresponding electron wave functions). The term $$\hat{H}_{D-MC}=L\left(\hat{B}^{\dag}_m\hat{D} +\hat{D}^{\dag}\hat{B}_m\right)$$ describes the interaction between the D/A molecule and the MC. Here, $ L$ is the transfer integral (energy of resonance dipole--dipole interaction) between the "donor" molecule and the particular structure element of MC. Usually, it can be assumed that $L\le J$. The term $\hat{H}_{D-MC}$ describes the excitation transition from the D/A molecule to MC. The mechanical oscillations of the MC (phonon ensemble) are described by $$\hat{H}_{ph}=\sum_q\hbar\omega_q \hat{b}^{\dag}_q\hat{b}_q\; ,$$ where $\hat{b}^{\dag}_q$ ($\hat{b}_q$) are the creation (annihilation) operators of the $q$--th phonon mode ($q$ is the phonon wave number). We supposed that the phonon ensemble is in thermodynamic equilibrium with the surrounding thermal bath at the temperature $T$. Finally, $$\hat{H}_{MC-ph}=\frac{1}{\sqrt{N}}\sum_{n,q}F_q \mathrm{e}^{iqnR_0}\hat{B}^{\dag}_n\hat{B}_n\left(\hat{b}_q+ \hat{b}^{\dag}_{-q}\right)$$ is the local interaction of the excitation placed on the $n$--th structure element of the MC with mechanical oscillations of the MC. In the case when the excitation interacts with optical phonons, the excitation--phonon interaction parameter has the form $F_q=\chi\sqrt{\hbar/(2M\omega_q)}$ (for non--dispersive optical phonons we have $F_q=F$), while in the case of interaction with acoustic phonons it has the form $F_q=2i\chi\sqrt{\hbar/(2M\omega_q)}\sin(qR_0)$. Here $R_0$ is the distance between adjacent structural elements of the molecular chain, $\omega_q=\omega_0\sin qR_0/2$ is dispersion law for phonons ($\omega_0=2\sqrt{\kappa/M}$, $\kappa$ is the "coefficient of elasticity" of the chain, $M$ is the mass of the molecular group of the chain) and $\chi$ is the excitation--phonon interaction constant \cite{AK,BI,ZdravkovicCevizovicND,PhysB2005Ivic, PhysB2009Cevizovic}.\\

Within the proposed model, the excitation interacts only with the mechanical oscillation of the structural elements of the MC. Consequently, we can remove the interaction term $\hat{H}_{MC-ph}$ from the Hamiltonian $\hat{H}$ by applying the Lang--Firsov unitary transformation $\hat{U}=\mathrm{e}^{-\sum_n\hat{B}^{\dag}_n\hat{B}_n\hat{S}_n}$ where $\hat{S}_n=\frac{1}{\sqrt{N}}\sum_q\frac{F_q}{\hbar\omega_q}\mathrm{e}^{iqnR_0}\left(\hat{b}_q-\hat{b}^{\dag}_{-q}\right)$ \cite{LF,YarkonyJCP,PhysB2009Cevizovic,ZdravkovicCevizovicND, CevizovicPRE}. After that, the transformed Hamiltonian $\hat{\bar{H}}=\hat{U}\hat{H}\hat{U}^{-1}$ is:

\begin{align}\label{HamE}
&\hat{\bar{H}}=
\mathcal{E}_D\hat{D}^{\dag}\hat{D}+\left(\mathcal{E}_0-\mathcal{E}_b\right)\sum_n\hat{B}^{\dag}_n\hat{B}_n\nonumber\\
&-J\sum_n\hat{B}^{\dag}_n\left(\hat{B}_{n+1}\mathrm{e}^{\hat{S}_{n+1}-\hat{S}_n}+\hat{B}_{n-1}\mathrm{e}^{\hat{S}_{n-1}-\hat{S}_n}\right)\\&
+L\left(\hat{B}^{\dag}_m\hat{D}\mathrm{e}^{-\hat{S}_m}+\hat{D}^{\dag}\hat{B}_m\mathrm{e}^{\hat{S}_m}\right)+\sum_q\hbar\omega_q\hat{b}^{\dag}_q\hat{b}_q\; .\nonumber
\end{align}

\noindent Here, $\hat{b}^{\dg}_q$ and $\hat{b}_q$ are creation and annihilation operators of new phonons, corresponding to the mechanical oscillations of MC structure elements around their new equilibrium positions. The operators $\hat{B}^{\dg}_n$ and $\hat{B}_n$ are creation and annihilation operators of new excitation, self--trapped on $n$--th structure element of MC and "dressed" by the cloud of new phonons. The parameter $\mathcal{E}_b=\frac{1}{N}\sum_q \frac{|F_q|^2}{\hbar\omega_q}$ is the so--called polaron "binding energy". As we have a single particle problem, the terms in Eq.(\ref{HamE}) that correspond to residual excitation--excitation interaction are neglected. Let us note that the "dressed" excitation and new phonons remain coupled in the third term of the Hamiltonian Eq.(\ref{HamE}), describing the migration of "dressed" excitation along MC (the term containing transfer integral $J$). Here, "dressed" excitation remains coupled with the phonon subsystem in an extremely non--linear way!\\

The further procedure is based on the application of the mean field approximation \cite{CevizovicCPB,CevizovicPRE,PhysB2005Ivic,PhysB2009Cevizovic,ZdravkovicCevizovicND}, that is, on the averaging of the particle subsystem over new phonons, which are in thermal equilibrium with the environment (characterized by the temperature $T$) of the macromolecule. In that case, we have $\hat{\mathcal{H}}_{exc}= \hat{\bar{H}}-\hat{\bar{H}}_{ph} -\hat{\bar{H}}_{rest}$, where $\hat{\bar{H}}_{rest}=\hat{\bar{H}}-\hat{\bar{H}}_{ph}-\left\langle\hat{\bar{H}} -\hat{\bar{H}}_{ph}\right\rangle_{ph}$ describes energy "fluctuations" of the particle subsystem 
around its mean energy. We consider these fluctuations small, so we ignore them. The notation $\left\langle\hat{A}\right\rangle_{ph}$ denotes the averaging of the operator $\hat{A}$ over the equilibrium ensemble of new phonons: $\left\langle\hat{A}\right\rangle_{ph}=\frac{1}{Z}\mathrm{Tr}\left\lbrace\mathrm{e} ^{-\beta\hat{\bar{H}}_{ph}}\hat{A}\right\rbrace$, $Z$ is the partition function of the phonon sub--ensemble, and $\beta=1/k_BT$ ($k_B$ is the Boltzmann constant). In this way, we obtain the Hamiltonian of the "dressed" excitation in the mean field approximation $\hat{\mathcal{H}}_{exc}=\left\lbrace\hat{\bar{H}} -\hat{\bar{H}}_{ph}\right\rbrace_{ph}$:

\begin{align}\label{HamEMF}
&\hat{\mathcal{H}}_{exc}=
\mathcal{E}_D\hat{D}^{\dag}\hat{D}+\left(\mathcal{E}_0-\mathcal{E}_b\right)\sum_n\hat{B}^{\dag}_n\hat{B}_n\nonumber\\
&-J\mathrm{e}^{-W_J(T)}\sum_n\hat{B}^{\dag}_n\left(\hat{B}_{n+1}+\hat{B}_{n-1}\right)+\\
&+L\mathrm{e}^{-W_L(T)}\left(\hat{B}^{\dag}_m\hat{D}+\hat{D}^{\dag}\hat{B}_m\right)\nonumber
\end{align}

\noindent where $W_J(T)=\frac{1}{N}\sum_q|f_q|^2\left(1-\cos(qR_0)\right) \text{coth}\left(\frac{\hbar\omega_q}{2k_BT}\right)$ and $W_L(T)=\frac{1}{2N}\sum_q |f_q|^2\text{coth}\left(\frac{\hbar\omega_q}{2k_BT}\right)$ are the renormalization factors of $J$ and $L$, respectively.

\subsection{Energy spectra of excitation in the D/A--MC structure}

To examine the influence of the D/A molecule on the energy spectra of "dressed" excitation located on the molecular bridge, we pass to the $k$ space using the transformation $\hat{B}_n=\frac{1}{\sqrt{N}}\sum_k\mathrm{e}^{-iknR_0}\hat{B}_k$. Here, quasi--particle wavenumber $k$ takes $N$ different values from the interval $kR_0\in\left[-\pi,\pi\right]$. We have:

\begin{align}\label{Hexck}
\hat{\mathcal{H}}_{exc}&=\mathcal{E}_D\hat{D}^{\dag}\hat{D}+\sum_k\mathcal{E}_k\hat{B}^{\dag}_k\hat{B}_k\\
&+\sum_k\left(\lambda^*_{m,k}\hat{B}^{\dag}_k\hat{D}+\lambda_{m,k}\hat{D}^{\dag}\hat{B}_k\right)\nonumber
\end{align}

\noindent where:

\begin{align}\label{EkLambda}
&\mathcal{E}_k=\mathcal{E}_0-\mathcal{E}_b-2J\mathrm{e}^{-W_J(T)}\cos(kR_0)\\
&\lambda_{m,k}=\frac{1}{\sqrt{N}}L\mathrm{e}^{-W_L(T)}\mathrm{e}^{-ikmR_0}\nonumber
\end{align}

\noindent Here, $\mathcal{E}_k$ is the energy band of the excitation, self--trapped on the MC. This expression is identical to the one, obtained for the small--polaron formed at the MC in the absence of the D/A molecule \cite{CevizovicCPB,CevizovicPRE, ZdravkovicCevizovicND}.

\begin{figure}[h]
	\begin{center}
		\includegraphics[height=4cm]{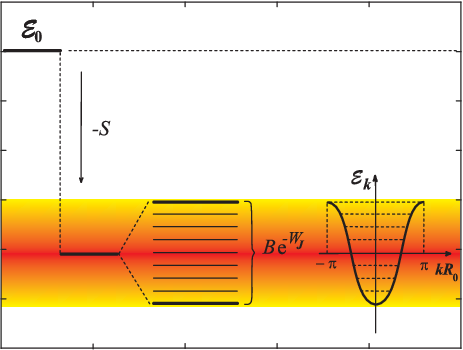}
	\end{center}
	\vskip -0.5cm
	\caption{The creation of a polaron energy band in 1D regular structure (ST excitation process).}\label{fig2}
\end{figure}

To find the energy spectrum of "dressed" excitation in the D/A--MC system, it is necessary to diagonalize the Hamiltonian Eq.(\ref{Hexck}). If we consider Eq.(\ref{Hexck}) as a quadratic form $$\hat{\mathcal{H}}_{exc}=\begin{bmatrix}\hat{D}^{\dag} & \hat{B}^{\dag}_1 & ...& \hat{B}^{\dag}_N\end{bmatrix}
\underbrace{\begin{bmatrix} \mathcal{E}_D & \lambda_1 & \lambda_2 & .... & \lambda_N\\
\lambda^*_1 & \mathcal{E}_1 & 0 & ... & 0\\
\lambda^*_2 & 0 & \mathcal{E}_2 & ... & 0\\ 
... & ... & ... & ... & ...\\
\lambda^*_N & 0 & 0 & ... & \mathcal{E}_N\end{bmatrix}}_{\hat{\mathbb{E}}} 
\begin{bmatrix}\hat{D} \\ \hat{B}_1 \\ ...\\ \hat{B}_N\end{bmatrix}$$ its complete diagonalization can be achieved by applying such unitary transformation that will diagonalize the matrix $\hat{\mathbb{E}}$ containing the coefficients from the Hamiltonian ({\ref{Hexck}}). Mathematically, this problem reduces to solving the eigenproblem of the coefficient matrix $\hat{\mathbb{E}}$, $$\hat{\mathbb{E}}\cdot\mathbb{X}=\mathcal{E}\cdot\mathbb{X}$$ which has non--trivial solutions provided that $\mathrm{det}\left(\mathbb{E}-\mathcal{E}\cdot\mathbb{I}\right)=0$ is satisfied. Here, symbol $\mathcal{E}$ represents the set of eigenvalues of the matrix $\hat{\mathbb{E}}$, and $\hat{\mathbb{X}}$ are corresponding eigenvectors. At the same time, the set of values of the parameter $\mathcal{E}$ determines the energy spectrum of the quasi--particle belonging to the entire D/A--MC structure. By solving the above determinant, we obtain the secular equation that determines the quasi--particle energy spectrum:

\begin{equation}\label{SekDav}
\mathcal{E}_D-\mathcal{E}-\sum_k\frac{|\lambda_k|^2}{\mathcal{E}_k-\mathcal{E}}=0
\end{equation}

\noindent where $|\lambda_k|^2=|\lambda|^2=\frac{1}{N}L^2\mathrm{e}^{-2W_L(T)}$ (it does not depend on $k$!). Because we are interested in stable states of the excitation in the D/A--MC structure, we will limit ourselves to the analysis of the lowest energy states obtained by solving Eq.(\ref{SekDav}). In addition, we will look for those solutions of Eq.(\ref{SekDav}) that satisfy the condition $\mathcal{E}_0-\mathcal{E} > 2J$. This condition is satisfied for quasi--particles that form the narrow energy band, and biomolecules, as a rule, belong to such a class of systems \cite{DavydovBQM}. The sum over the $k$ can be easily calculated if we replace it with the integral $$\frac{1}{N}\sum^{\pi/R_0}_{k=-\pi/R_0} A_k\rightarrow\underbrace{\frac{R_0}{2\pi} \int^{\pi/R_0}_{-\pi/R_0}A(k)dk}_{x=R_0k}=\frac{1}{2\pi}\int^{\pi}_{-\pi}A(x)dx\;.$$ Besides, it is useful to present the obtained expressions by the set of non--dimensional system parameters: (polaron--phonon) coupling constant $S=\mathcal{E}_b/\hbar\omega_0$, adiabatic parameter $B=2J/\hbar\omega_0$, the normalized energy of the excitation in the D/A--MC structure $\bar{\mathcal{E}}=\mathcal{E}/ \hbar\omega_0$, normalized energy level of the excitation on the D/A molecule  $\bar{\mathcal{E}}_D=\mathcal{E}_D/\hbar\omega_0$, and transfer parameter $\gamma=L/J$. The parameter $L$ provides information about the probability of excitation delocalization from the D/A molecule to the nearest structure element of the MC, while parameter $J$ provides information about excitation delocalization from one structure element to the neighboring one along the MC. Thus, the relative ratio of these two parameters tells us how much easier (or harder) the excitation can migrate along the MC, compared to its delocalization from one molecule to another. After introducing the set of non--dimensional parameters and performing the above--mentioned integration, we obtain:

\begin{equation}\label{EdispSB}
\bar{\mathcal{E}}_D-\bar{\mathcal{E}}-\frac{\gamma^2B^2\mathrm{e}^{-2S\coth(1/2\tau)}}{4\sqrt{(\bar{\mathcal{E}}_0-S-\bar{\mathcal{E}})^2-B^2\mathrm{e}^{-2S\coth(1/2\tau)}}}=0\;.
\end{equation}

\noindent Here, $\tau=k_BT/\hbar\omega_0$ is the normalized temperature. The obtained equation determines the lowest energy of "dressed" excitation belonging to the D/A--MC structure. This equation is the main result of our analysis. It allows us to analyze how the properties of injected excitation depends on basic energy parameters of the structure and the environment temperature. Let's notice here that the solutions of Eq.(\ref{EdispSB}) must satisfy the condition:

\begin{equation}\label{ConditionE}
(\bar{\mathcal{E}}_0-S-\bar{\mathcal{E}})^2-B^2\mathrm{e}^{-2S\coth(1/2\tau)}> 0\;.
\end{equation}

\noindent This condition excludes all values of the $\bar{\mathcal{E}}$ belonging to the interval $\mathcal{E}_{UP}\leq \bar{\mathcal{E}}\leq \mathcal{E}_{LO}$. Here, $\mathcal{E}_{LO}=\bar{\mathcal{E}}_0-S-B\mathrm{e}^{-S\coth(1/2\tau)}$ and $\mathcal{E}_{UP}=\bar{\mathcal{E}}_0-S+ B\mathrm{e}^{-S\coth(1/2\tau)}$ and they correspond to the borders of the small--polaron energy band, formed due to the excitation self--trapping at the MC, but in the absence of the D/A molecule. Let us note that the ambient temperature affects the excitation energy $\bar{\mathcal{E}}$ only through the exponential factor $\mathrm{e}^{-2S\coth(1/2\tau)}$, which reduces the adiabatic parameter $B$: $B(\tau)=B\mathrm{e}^{-2S\coth(1/2\tau)}$. In principle, the increasing of the ambient temperature $\tau$ leads to a rapid reduction of the value of the parameter $B$. This fact enable us to examine the temperature influence on the excitation energy only through the influence of the adiabatic parameter $B$.\\

To compare the obtained results with the non--adiabatic polaron picture, it is most convenient to base further analysis on the study of the dependence of $\mathcal{E}(S)$. At the same time, $B$, $\gamma$ and $\tau$ can be (mathematically) treated as parameters \cite{CevizovicCPB,CevizovicPRE,PhysB2009Cevizovic,BI}. The theoretical analysis of the proposed model requires the numerical solution of the obtained results. Further, this requires a numerical estimation of the basic parameters of the considered system. Since our main goal is to analyze the possibility of stable excitation transfer from the D/A molecule to the MC in biological processes under the assumption that the excitation to MC forms a non--adiabatic polaron, for the values of the basic system parameters we will adopt widely used data from the relevant literature. In the $\alpha$--helix protein MC, the value for the transfer--integral $J$ between two peptide units along the covalent bond is about $J_{cov}\approx -12.4\;\text{cm}^{-1}$, while in the case of the hydrogen bonded peptide units, it is $J_{hb}\approx 7.8\;\text{cm}^{-1}$ \cite{ScottPR,ScottPRA,Nevskaya}. At the same time, the mass of the structure element in $\alpha$--helix MC is about $M\approx 5.7\cdot10^{-25}\;\text{kg}$ \cite{ScottPR,ScottPRA}. The coefficient of the elasticity of the covalent bond $\kappa$ varies between 45 N/m and 75 N/m, while for the hydrogen bond it ranges from 13 N/m to 20 N/m \cite{FalvoPouthier,ScottPR,HennigPRB}. These values indicate that the typical phonon frequency is of the order of magnitude $\omega_0\approx 10^{13}\;\text{s}^{-1}$. Finally, excitation--phonon interaction constant $\chi$ ranges from 45 N/m to 75 N/m in the case of the covalent bond \cite{HennigPRB}, and from 13 N/m to 20 N/m in the case of the hydrogen bond \cite{ScottPR,FalvoPouthier, HennigPRB}. The typical solution of Eq.(\ref{EdispSB}) is presented in Fig.(\ref{fig3}):

\begin{figure}[h]
\begin{center}
\includegraphics[height=5cm]{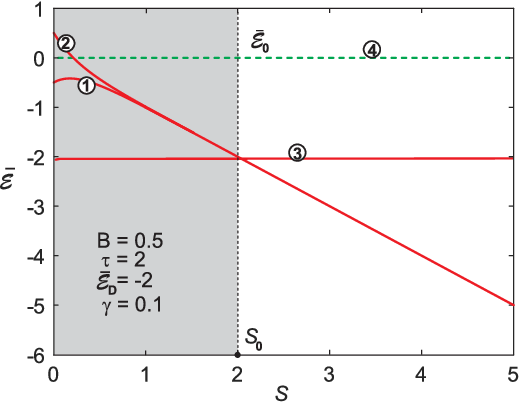}
\end{center}
\vskip -0.5cm
\caption{Typical dependence $\bar{\mathcal{E}}(S)$ for the fixed values of $\bar{\mathcal{E}}_0$, $\bar{\mathcal{E}}_D$, $\tau$, $B$, and $\gamma$. Line "4" is the excitation energy  $\mathcal{E}_0$ of the "bare" excitation on the MC. The $S_0$ is the intersection point of the branches "1" and "3".}\label{fig3}
\end{figure}

For the fixed values $S$, $B$, $\gamma$, $\tau$, $\bar{\mathcal{E}}_D$ and $\bar{\mathcal{E}}_0$, Eq.(\ref{EdispSB}) has three solutions. Consequently, the dependence $\bar{\mathcal{E}}(S)$ can be represented by a curve with three branches (denoted by numbers 1, 2, and 3 in Fig.(\ref{fig3})). In principle, these curves can provide information about the excitation state. To obtain a more detailed insight into the nature of these branches, let us consider the energy spectrum of $\hat{\mathcal{H}}_{exc}$ when $\gamma\ll 1$. In that case, the last term of the Hamiltonian Eq.(\ref{Hexck}) can be treated as a perturbation $$\hat{H}_{perturb.}= \sum_k\left(\lambda^*_{m,k} \hat{B}^{\dg}_k\hat{D}+\lambda_{m,k}\hat{B}_k \hat{D}^{\dg}\right).$$ The energy spectrum of the unperturbed part of the Hamiltonian Eq.(\ref{Hexck}) is determined by the energy level $\bar{\mathcal{E}}_D$ (corresponding to the quantum state of the single excitation localized on the D/A molecule $\left|1_D\right\rangle= \hat{D}^{\dg}\left|0\right\rangle$) and the energy spectra $\bar{\mathcal{E}}_k$ (corresponding to the states $\left|1_{k}\right\rangle=\hat{B}^{\dg}_{k} \left|0\right\rangle$ of self--trapped excitation localized on MC) which form an energy band due to the assumed translation invariance of the molecule. Due to the presence of perturbation, the energies of these states in the second order of the perturbation theory become (for the $N\ll 1$):

\begin{align}\label{EDEkPERTURB}
&\bar{\mathcal{E}}^{(pert)}_D=\bar{\mathcal{E}}_D-\frac{\gamma^2B}{4}\tilde{I}_D(2)\\
&\bar{\mathcal{E}}^{(pert)}_k=\bar{\mathcal{E}}_k=\bar{\mathcal{E}}_0-S-B\mathrm{e}^{-S\mathrm{coth}(1/2\tau)}\cos(kR_0)\;,\nonumber
\end{align}

\noindent where $I_D(2)=\frac{1}{N}\sum_k\frac{1}{p-\cos(kR_0)}$, $p=\frac{\bar{\mathcal{E}}_0-S-\bar{\mathcal{E}}_D}{B}\mathrm{e}^{W_J(\tau)}$. At the same time, the vectors corresponding to these (perturbed) energies differ slightly from the initial (unperturbed) ones. This means that the perturbed state originating from the state $\left|1_D\right\rangle$ (for example) is still very close to the initial one. The quasi--particle whose distribution of the probability of finding has a dominant value in the vicinity of the D/A molecule remains localized on the D/A molecule even after the appearance of the perturbation (that is, after the appearance of the interaction with MC). Therefore, we can consider that the excitation that was initially located on the D/A molecule, after D/A molecule comes into contact with the MC slightly changes excitation energy, but it still remains dominantly localized on the D/A molecule.
The same conclusion is valid in the case of states corresponding to excitation initially localized on the MC. Of course, when the interaction of the D/A subsystem with the MC becomes large enough, this picture ceases to be applicable. Then the entire structure represents a unique quantum system, and the excitation belongs to the structure as a whole. But even in this case, we will call the states of a quasi--particle whose energy is close to $\bar{\mathcal{E}}_D$ as states of a particle "localized" on D/A molecule, and those states close to energies $\bar{\mathcal{E}}_k$ as states of "dressed" excitation "localized" on MC.\\

\begin{figure}[h]
	\begin{center}
		\includegraphics[height=5cm]{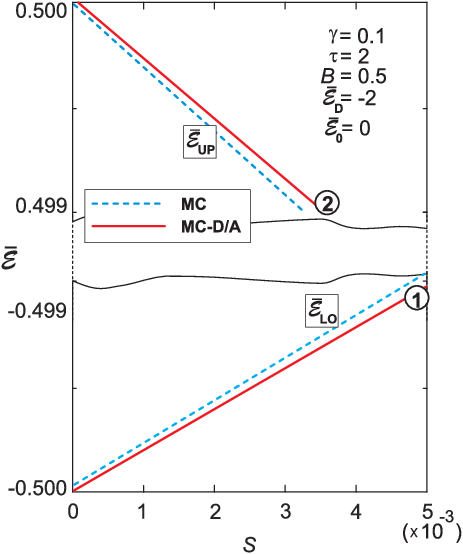}
		\includegraphics[height=5cm]{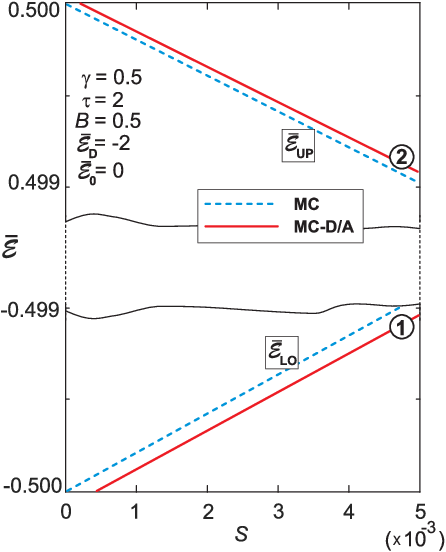}
	\end{center}
	\vskip -0.5cm
	\caption{The dependence $\bar{\mathcal{E}}(S)$ in the area $S\ll 1$, for different values of $\gamma$. Here, the full line presents the result obtained by solving Eq.(\ref{EdispSB}). By the dashed line, we presented $\mathcal{E}_{UP}$ and $\mathcal{E}_{LO}$ (the upper and the lower border of the small--polaron band in the MC), in the absence of the A/D subsystem.}\label{fig4}
\end{figure}

As it can be noticed from Eq.(\ref{EDEkPERTURB}), for $\gamma\ll 1$ the presence of the D/A molecule does not affect the energy levels $\bar{\mathcal{E}}^{(pert)}_k$. Here, $\bar{\mathcal{E}}^{(pert)}_{k=-\pi}$ coincide with $\mathcal{E}_{UP}$ and $\bar{\mathcal{E}}^{(pert)}_{k=0}$ coincide with $\mathcal{E}_{LO}$. The difference between the dependence $\bar{\mathcal{E}}(S)$ obtained by the perturbation method and the one, obtained by solving Eq.(\ref{EdispSB}) is shown in Fig.({\ref{fig4}}). As we can see, the branches "1" and "2" are quite close to those curves, corresponding to the upper and the lower borders of the energy band of non--adiabatic polaron quasi--particle, emerging by self--trapping of excitation in the MC  {\bf in the absence of D/A molecule}.
For all values of the system parameters, branch "1" has slightly larger values than $\mathcal{E}_{UP}$, while branch "2" has slightly lower values than $\mathcal{E}_{LO}$. As a consequence, the area bordered by curves $\mathcal{E}_{LO}$ and $\mathcal{E}_{UP}$ lies inside the area bordered by branches "1" and "2". So, the states belonging to these branches have the physical significance because they satisfy the condition Eq.(\ref{ConditionE}). As we can see from Fig(\ref{fig3}), both branches show typical dependence of the energy of the small--polaron band on the coupling constant $S$, including for large values of $S$, where the dependence $\mathcal{E}(S)$ has the form $\mathcal{E}_{SP}\sim -S$. Like "standard" non--adiabatic polaron, the "dressed" excitation located on the MC has the properties of the fully dressed, hardly mobile quasi--particle. That is the reason why the branches "1" and "2" can be identified as the upper and the lower borders of the small--polaron energy band, for polaron self--trapped on MC, even in the case when $\gamma\sim 1$.

The result obtained here is consistent with our expectations since the MC is usually a much larger system than the D/A molecule and the interaction between these two subsystems is much weaker than that, between the neighboring structural elements of the MC itself. If the D/A molecule were larger it could interact with the MC not only locally. This would also cause the interaction of the D/A molecule with the mechanical oscillations of the MC, which could significantly change the properties of the small--polaron formed due to the excitation--phonon interaction.\\

\begin{figure}[h]
	\begin{center}
		\includegraphics[height=5cm]{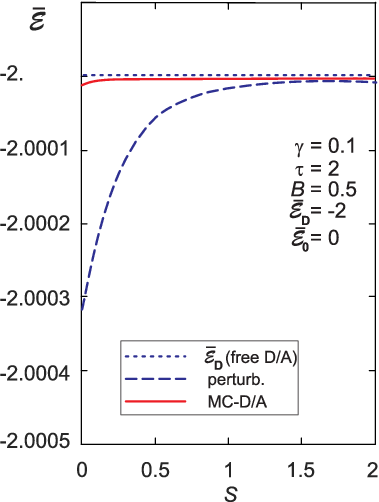}
		\includegraphics[height=5cm]{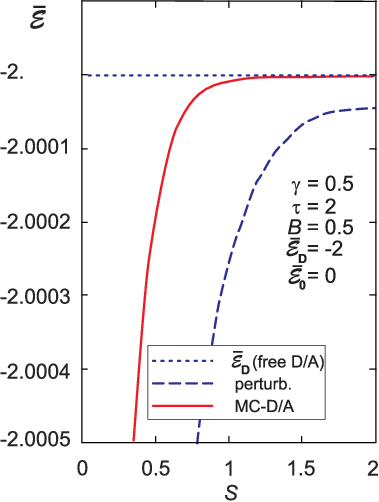}
	\end{center}
	\vskip -0.5cm
	\caption{The dependence $\bar{\mathcal{E}}(S)$ obtained by solving of Eq.(\ref{EdispSB}) and that, predicted by perturbation theory.}\label{fig5}
\end{figure}

At the same time, the energy level of the excitation located on the D/A molecule is slightly reduced. The magnitude of this reduction depends on system parameters and ambient temperature. To see the impact of this correction, let's plot the $\bar{\mathcal{E}}^{(pert)}_D(S)$ dependence predicted by Eq.(\ref{EDEkPERTURB}). From Fig.(\ref{fig5}) we can see that $\bar{\mathcal{E}}(S)$ obtained by perturbation calculation approximates branch "3" well for large values of the parameter $S$. The difference is significant for small values of this parameter only. With increasing $\gamma$, the difference between the curve obtained by perturbation theory and our model becomes larger (right pane). As we can remark, the energies presented by branch "3" are close to the energy of the excitation, localized on the D/A molecule in the absence of the MC (free D/A). Therefore, in the case when $\gamma\sim 1$ the states corresponding to the points from the branch "3" are the states of the excitation, localized on the D/A molecule.\\

Since they represent the lowest values of the "dressed" excitation energy, the branches "1" and "3" play the most important role in the analysis of polaron stability. Here, the branch 1 corresponds to the bottom of the small--polaron band states, in the case when the "dressed" excitation is located on MC. At the same time, the branch "3" corresponds to the state of the excitation, localized on the D/A molecule.\\

Now we can do the basic interpretation of the graph in Fig.(\ref{fig3}). According to obtained results, in the parameter space there are two basic regimes: the first one, where the excitation is localized on the D/A molecule, and the second one, where it migrates to the MC and form the "dressed" quasi--particle. These two regimes are separated by the intersection point $S_0$. In the parameter space, for all $S<S_0$, energy values represented by curve "3" are the most energetically stable solutions i.e. the excitation remains to stay on the D/A molecule. For the values $S>S_0$, we find that the energetically most favorable state corresponds to the branch "1". Here, the energy of "dressed" excitation linearly decreases with the increase of $S$: $\mathcal{E}\sim -S$. On the other side, according to the standard polaron theory \cite{ZdravkovicCevizovicND, EminPRB, Rashba, AK}, such dependence is characteristic for non--adiabatic polaron quasiparticle. Therefore, we expect that for these values of the system parameters, the excitation initially located on the D/A molecule will pass to the MC and (due to the interaction with the phonons of the MC) forms a hardly mobile but quite stable non--adiabatic polaron state.



\section{The impact of the D/A presence on the non--adiabatic polaron state}

Let us now examine the impact of D/A presence on the excitation state in the D/A--MC structure. Here, we are primarily interested in what conditions must be satisfied, so that the D/A molecule behaves as a donor, that is, as an acceptor system. In general, for the D/A molecule to be a donor, the condition $\mathcal{E}_D>\mathcal{E}_0$ must be fulfilled. Due to the polaronic effect on the MC, this condition takes the form $\mathcal{E}_D> \mathcal{E}_0-\mathcal{E}_b$, or in terms of the normalized parameters, we need $\bar{\mathcal{E}}_D>\bar{\mathcal{E}}_0-S$. However, due to the interaction between the D/A molecule and the MC, the condition becomes much more complex and comes down to examining the mutual position of the bottom of the polaron energy band and the energy level of the excitation localized on the D/A molecule. In other words, it comes down to examining the relative position of branches "1" and "3", obtained by solving Eq.(\ref{EdispSB}). This can be done by checking the influence of parameters that appear within the framework of our model as a direct consequence of the existence of the D/A molecule, such as $\gamma$ and $\mathcal{E}_D$. Indirectly, the presence of the D/A molecule can affect the properties of the "dressed" excitation by changing the properties of the MC, that is, by changing the values of those system parameters that are the primary characteristic of the MC. Such parameters are $S$ and $B$.

\subsection{The influence of $\gamma$}

Here, we assumed that the transfer parameter $\gamma$ can take values from $\gamma\ll 1$ to $\gamma\sim 1$, that is, we suppose that $L$ and $J$ can be of the same order of magnitude.

\begin{figure}[h]
	\begin{center}
		\includegraphics[height=5cm]{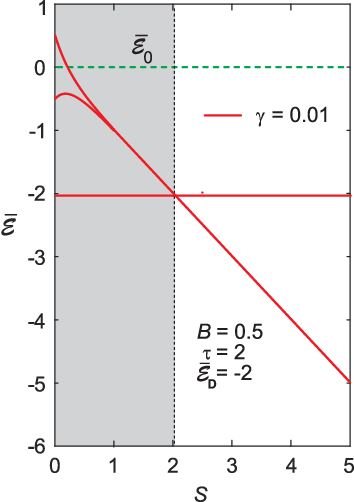}
		\includegraphics[height=5cm]{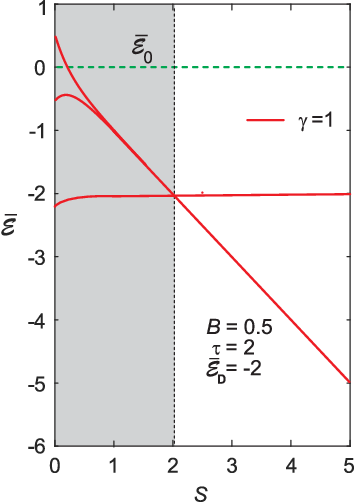}
	\end{center}
	\vskip -0.5cm
	\caption{The dependence $\bar{\mathcal{E}}(S)$, for different values of the $\gamma$.}\label{fig6}
\end{figure}

\begin{figure}[h]
	\begin{center}
		\includegraphics[height=6cm]{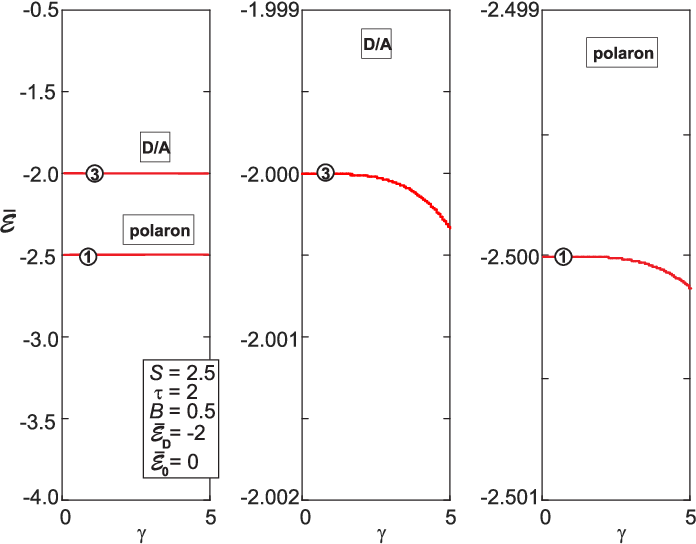}
	\end{center}
	\vskip -0.5cm
	\caption{The dependence $\bar{\mathcal{E}}(\gamma)$, obtained by the solving of Eq.(\ref{EdispSB}) in strong coupling limit ($S>1$). Left pane: the dependence of the bottom of the polaron energy band (branch "1") and excitation energy level on D/A molecule (branch "3") on $\gamma$ can not be seen at the chosen scale. To show this, it is necessary to choose the suitable scale for each branch separately (the middle and the right panes). Middle pane: the dependence of the branch "3" on the $\gamma$. Right pane: the dependence of the branch "1" on the $\gamma$. The values of the fixed parameters were chosen so that the graphs show the area to the right of the point $S_0$ (the area $\bar{\mathcal{E}}(D/A)>\bar{\mathcal{E}}(polaron)$).}\label{fig7}
\end{figure}

In Fig.(\ref{fig6}), we presented the dependence of the energy of "dressed" excitation $\bar{\mathcal{E}}(S)$ on the polaron--phonon coupling strength $S$ for several values of $\gamma$. Numerical analysis of Eq.((\ref{EdispSB}) shows that the increasing of $\gamma$ slightly increases the distance between branches "1" and "2" (spreads the polaron energy band). With the increase of $\gamma$, numerical values of the $\bar{\mathcal{E}}$ corresponding to the bottom of the polaron energy band (the branch "1") attain smaller values. At the same time, numerical values that correspond to the top of the polaron energy band (the branch "2") attain slightly larger values. As a consequence, the width of the polaron energy band increases slightly with increasing $\gamma$. Despite that, for $S>1$ the width of the polaron energy band remains quite narrow, and the influence of the presence of the D/A molecule here is practically negligible. The spreading of the branches "1" and "2" is very small, and at the scale in Fig.(\ref{fig6}) it is not visible, even for $S\ll 1$. A better insight into this can be gained from the right pane of Fig.(\ref{fig7}), where the dependence of the branch "1" on $\gamma$ is presented on a more convenient scale.\\

At the same time, the values of $\bar{\mathcal{E}}$ from branch "3" decrease with the increase of $\gamma$. For large values of $S$, the displacement of the energies from the branch "3" is significantly smaller, but it is more significant than the displacements of the branches "1" and "2", even for large values of $\gamma$.

As we can see, the change of the $\gamma$ practically does not affect polaron states on the MC, but it influences the excitation energy level on the D/A molecule. This influence is more significant in weak coupling limit ($S\ll 1$), where all branches of the excitation energy spectrum are distorted. In the strong coupling limit, the change of the $\gamma$ practically does not affect the excitation energy level on the D/A molecule nor the energy of the polaron state on MC. As a consequence, the position of $S_0$ practically is not affected.

\subsection{The influence of $\mathcal{E}_D$}

Here, we examine the influence of the difference between the excitation energy on the D/A molecule $\mathcal{E}_D$ and the the energy corresponding to the non--trapped state of excitation in the MC $\mathcal{E}_0$ on the properties of the single excitation in the D/A--MC structure.

\begin{figure}[h]
	\begin{center}
		\includegraphics[height=6cm]{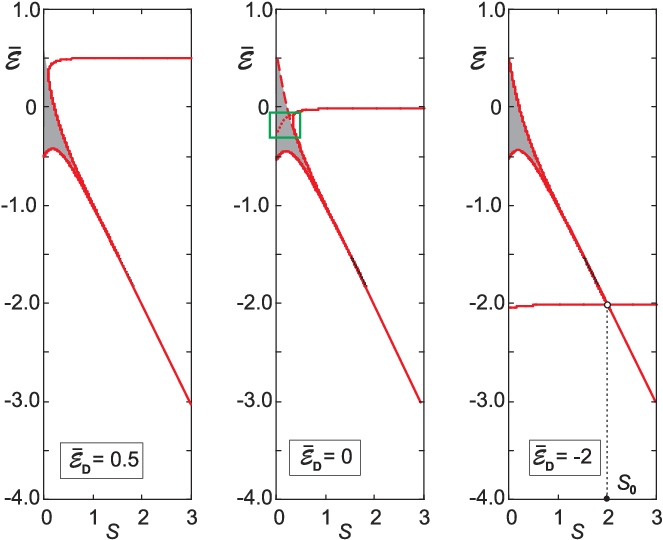}
	\end{center}
	\vskip -0.5cm
	\caption{The dependence $\bar{\mathcal{E}}(S)$, for $\bar{\mathcal{E}}_0=0$, $\gamma=1$, $B=0.5$, $\tau=2$, for three values of $\mathcal{E}_D$.}\label{fig8}
\end{figure}

In the graphic Fig.(\ref{fig8}), we have shown the dependence of $\bar{\mathcal{E}}(S)$ for several values of $\bar{\mathcal{E}}_D$. The first we can notice is that the position of the polaron energy band (the branches "1" and "2") practically does not depend on the difference $\bar{\mathcal{E}}_D-\bar{\mathcal{E}}_0$. As can be seen, in the case $\bar{\mathcal{E}}_D-\bar{\mathcal{E}}_0 >0$, the polaron state is energetically more favorable than the state of the excitation localized on the D/A molecule for all values of the parameter $S$ (left pane on Fig.(\ref{fig8})). Physically, this means that the excitation can perform the transition from D/A to MC, regardless of the coupling strength of the excitation with the phonon subsystem of the MC. The only condition that should be satisfied is that the excitation can form the non--adiabatic polaron state at MC. 
This means that $S$ is large enough to form a quantum well where excitation can be "captured" \cite{AK,CevizovicCPB,CevizovicPRE}. The transfer parameter $\gamma$ has no significant influence in this case, except for the extremely large values of $\gamma$ and small values of $S$. In that case, branch "3" can be so distorted that the exciton states localized on the D/A molecule become energetically more favorable than the polaron state on the MC. A similar situation is observed for the case $\bar{\mathcal{E}}_D =\bar{\mathcal{E}}_0$. 
Let us note that in this case the energy level of the excitation localized on the D/A molecule can be "nested" in the polaron energy band (rectangled area on the middle pane of Fig.(\ref{fig8})). For these values of the system parameters, the excitation is in such a quantum state that it is delocalized to the entire D/A--MC structure, and it easily migrates from one subsystem to another and vice versa.\\

The situation is quite different when $\bar{\mathcal{E}}_D- \bar{\mathcal{E}}_0<0$. In that case, there is a threshold value $S_0$ for the parameter $S$, below which the excitation remains on the D/A molecule, although the condition for the formation of the non--adiabatic polaron state on the MC is satisfied. 
But, for $S>S_0$  the D/A molecule can inject the excitation on the molecular chain. The effect is more pronounced for the strong coupling limit. At the same time, the larger difference $\bar{\mathcal{E}}_D-\bar{\mathcal{E}}_0$ implies the larger critical value $S_0$ (that is, the intersection point of the curves "1" and "3" moves toward larger values).


\subsection{The influence of $B$ and $\tau$}

Finally, we analyzed the influence of the system temperature $\tau$ and the adiabatic parameter $B$ on excitation energy in the D/A--MC structure.

\begin{figure}[h]
	\begin{center}
		\includegraphics[height=5cm]{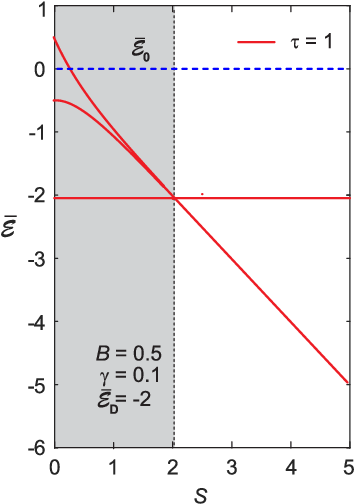}
		\includegraphics[height=5cm]{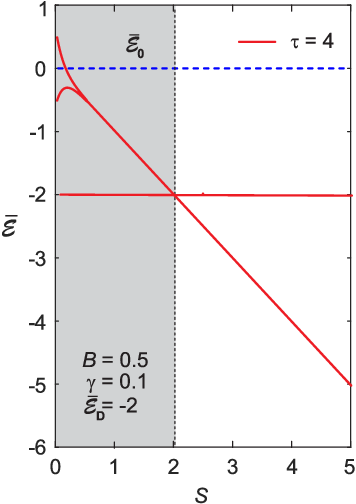}
	\end{center}
	\vskip -0.5cm
	\caption{The dependence $\bar{\mathcal{E}}(S)$, for different values of $\tau$. The value $\tau=4$ corresponds to the room temperature.}\label{fig9}
\end{figure}

In Fig.(\ref{fig9}) it was presented the dependence $\mathcal{E}(S)$ for different values of the normalized temperature $\tau$. Similarly to previous cases, the most remarkable changes in the excitation energy spectra occur in the area $S\ll 1$. According to obtained results, the increasing of the system temperature $\tau$ brings the curves "1" and "2" closer, that is, narrows the energy band of "dressed" excitation. As we mentioned, from Eq.(\ref{EdispSB}) we see that the temperature affects the energy of a small--polaron only through the normalization factor $\mathrm{e}^{-S\coth(1/2\tau)}$ of the $B$. Namely, as the temperature increases, the value of $B$ is scaled, and it rapidly decreases. \\

\begin{figure}[h]
	\begin{center}
		\includegraphics[height=5cm]{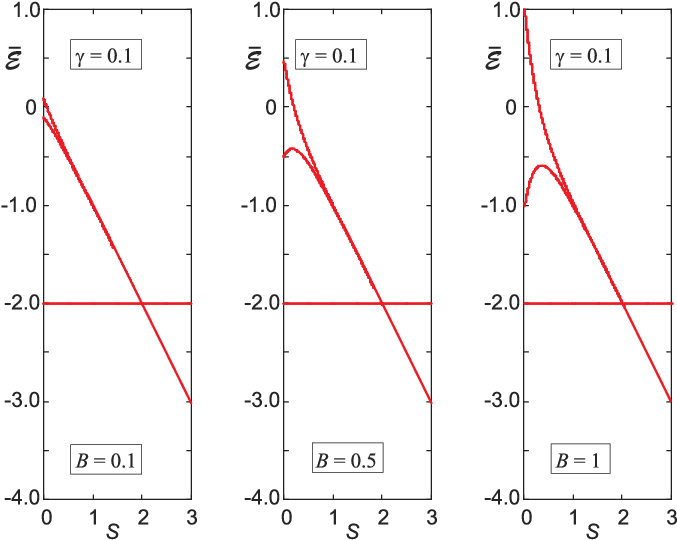}\\
		\includegraphics[height=5cm]{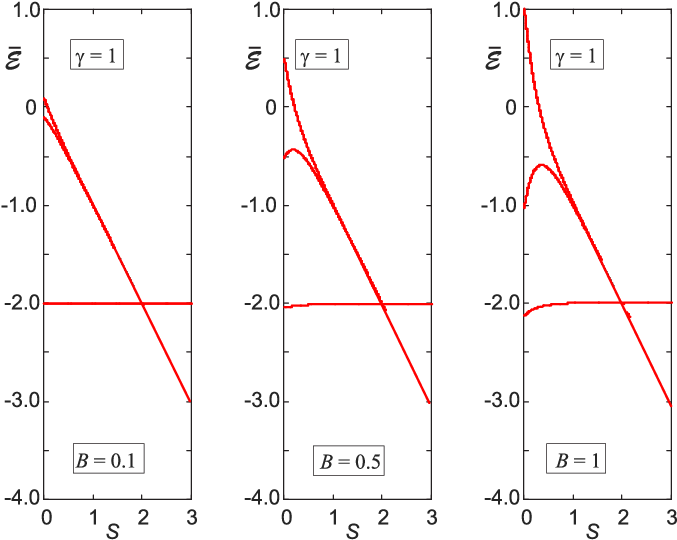}
	\end{center}
	\vskip -0.5cm
	\caption{The dependence $\bar{\mathcal{E}}(S)$, for $\tau=2$, $\bar{\mathcal{E}}_0=0$, $\bar{\mathcal{E}}_D=-2$, for various values of  $B$.}\label{fig10}
\end{figure}

Let us now examine the influence of the adiabatic parameter $B$ on the energy spectrum of the ST exciton state in the D/A--MC system. From Fig.(\ref{fig10}) one can remark that the increasing of $B$ affects all branches of excitation spectra in the weak couppling and the adiabatic limits ($S\ll 1$ and $B\sim 1$). But in the non--adiabatic and the strong coupling limits, the energy spectrum of the excitation shows, on the one hand, "strict" properties of the non--adiabatic polaron, and on the other hand, the properties of the excitation localized on the D/A molecule (depending on whether the values of the system parameters correspond to the area below or above of $S_0$).

From the graphics presented in (Fig.(\ref{fig9}) and Fig.(\ref{fig10})), we can remark that the decrease of $B$ (or, equivalently, the increase of $\tau$) does not affect the condition required for excitation transfer from D/A molecule on MC (and vice versa), that is, to the position of the $S_0$. It determines the width of the small--polaron energy band and consequently, its effective mass when excitation migrates on MC and forms the small--polaron state.

\section{Conclusion}

Let us now briefly summarize the main results of the presented paper. We have shown that the excitation injected into the molecular chain by the "donor" molecule can form (due to the interaction with the phonons of the MC) a "dressed" quasi--particle, the properties of which correspond to non--adiabatic polaron. On the other hand, the "acceptor" molecule can capture the excitation from the MC and destroy the polaron state. Whether the D/A molecule will behave as a "donor" or an "acceptor" depends on the values of the basic energy parameters of the structure. First of all, on the values of $\mathcal{E}_D$, $\mathcal{E}_0$, and $\mathcal{E}$. Structures whose system parameters correspond to the area from the left of $S_0$ are typical "acceptor" systems, and those whose parameters are to the right of $S_0$ represent typical "donor" structures (of course, the condition $S\gg 1$ and $B\ll 1$ must be satisfied).\\

According to the proposed model, the presence of the D/A molecule does not affect the state of the non--adiabatic polaron localized on the MC. On the other hand, the MC can significantly change the excitation state of the D/A molecule. Its influence is most pronounced for large values of $\gamma$ and in the weak interaction limit. The influence of the ambient temperature is reflected primarily in the reduction of the adiabatic parameter $B$. As the temperature increases, the system becomes more non--adiabatic, and the polaron energy band narrows. As a consequence, the inertness of the polaron increases.\\

Our conclusions qualitatively agree with the results provided by applying soliton models (we have in mind the results obtained by applying the Davydov soliton model). \cite{DavydovBQM,DavydovPSS,DavydovSMS, DavydovZETF,BrizhikPRE2014}. Moreover, they are expected because the D/A molecule is significantly smaller compared to the MC and interacts with it only locally. In the proposed model, this fact reflects in neglecting the influence of the D/A molecule on the mechanical oscillations of the MC. 
When the D/A molecule is large, its presence could change the exciton--phonon interaction constant $F_q$ on the MC and indirectly affect the polaron properties.
There is another interesting mechanism that can potentially change the phonon spectra of the MC and thus influence the properties of the energy spectra of the D/A--MC system. Namely, formed polaron can reversibly change the phonon spectrum of the medium in which the polaron forms. Such "feedback" effect of polaron has been studied in both adiabatic \cite{KalosakasPRB} and non--adiabatic limits \cite{PhysB2005Ivic}. According to the results of these studies, the "feedback" effect leads to the hardening of the phonon modes in non--adiabatic limit. As a consequence, all the branches of the energy spectrum of the D/A--MC system can be changed.\\ 


In addition, we mentioned that the values of the basic parameters of biomolecules are not known exactly. The discussion about whether they belong to the adiabatic or non--adiabatic limes, as well as whether the interaction of the excitation with the mechanical oscillation of the MC belongs to the limits of the strong, the medium, or even the weak exciton--phonon interaction continues until now. Therefore, even within the non--adiabatic limit, several models were proposed to explain some effects that did not fit into the "standard" non--adiabatic polaron model. One class of such models are so--called partial dressing models \cite{BI,CevizovicCPB,CevizovicPRE}. These models are based on a variational approach and the assumption that phonon modes participate only partially in excitation dressing and the polaron formation. According to the partial dressing models, there are two polaron solutions in polypeptide chains. The first one corresponds to a weakly dressed, almost free excitation, while the second one corresponds to a heavily dressed "standard" non--adiabatic polaron. The transition between these two states in the parameter space occurs abruptly. The qualitatively same behavior was undoubtedly confirmed employing the numerically exact diagonalization of the Holstein model where the discontinuous transition from free electron to immobile small polaron has been predicted \cite{AlvermanPRB} and in the investigation of the dimensionality effect on large to small polaron crossover in Holstein's model \cite{HammTsironis}. At the same time, the boundary in the parameter space that separates these two solutions depends on the environment temperature \cite{CevizovicCPB}. In that sense, it would be interesting to consider how the presence of the D/A molecule affects these two solutions, especially the values of the system parameters that are characteristic of the boundary region separating the mentioned solutions. However, this goes beyond the scope of this paper and will be the subject of our further research.

\section{Acknowledgment}

This work was supported by the Ministry of Science, Technological Development, and Innovation of the Republic of Serbia through the Project contract No 451-03-47/2023-01/200017. The authors would like to acknowledge the contribution of the COST Action CA21169, supported by COST(European Cooperation in Science and Technology).

\end{document}